\documentstyle[referee]{l-aa}

\begin{document}
\def\be{\begin{equation}}
\def\ee{\end{equation}}
\def\bea{\begin{eqnarray}}
\def\eea{\end{eqnarray}}
\def\nn{\nonumber}
\def\th{\theta}
\def\ph{\phi}
\def\lt{\left}
\def\rt{\right}
\input epsf.tex


   \thesaurus{12         
              (12.07.1;  
               02.18.8;  
               02.07.1;   
               02.02.1)  
               } 
   \title{Role of the scalar field in gravitational lensing}


   \author{K. S. Virbhadra\inst{1-2}
     \and  D. Narasimha\inst{1}
     \and  S. M. Chitre\inst{1}
          }

   \offprints{K. S. Virbhadra}

   \institute{ 
              Theoretical Astrophysics Group\\
              Tata Institute of Fundamental Research\\
              Homi Bhabha Road, Colaba\\
              Mumbai (Bombay) 400005, India\\
              \and
              Department of Applied Mathematics\\
              University of Zululand\\
              Private Bag X1001\\
              Kwa-Dlangezwa 3886\\
              South Africa
               }

  \date{Received 30 January 1998 / Accepted 30 April 1998}
  \date{}

   \maketitle

\begin{abstract}
A  static and circularly symmetric lens characterized by mass and scalar 
charge parameters is constructed.  For the small values of the scalar
charge to the mass ratio, the gravitational lensing is qualitatively 
similar to the case of the Schwarzschild lens; however, for  large values
of this ratio the lensing characteristics are significantly  different.
The main features  are the existence of two or nil Einstein ring(s)
and a radial critical curve, formation of two or four images and possibility
of detecting three images near the lens for sources located at relatively
large angular positions. Such a novel  lens may also be treated as a naked
singularity lens.

\end{abstract}

%

\section{Introduction}
The deflection of light in a gravitational field predicted by the general 
theory of relativity witnessed experimental verification in 1919. 
 The spectacular phenomena resulting from the deflection of
light in a gravitational field are  referred to as the
gravitational lensing (GL).  Chwolson, in 1924, pointed out that if a
background star (source), a deflector (gravitational lens) and an observer 
are perfectly aligned, a ring-shaped image of the source 
centered on the deflector would appear (see in Schneider et al. 1992).
This is, however, usually called as the `Einstein Ring'. Einstein (1936) obtained  the
apparent luminosities of the images of a star (source) due to a
foreground star (gravitational lens). He mentioned that if the source,
the lens and the observer are sufficiently aligned, the images will be highly
magnified. However, arguing that the  angular  image splitting
caused by a stellar-mass lens is too small to be resolved by an
optical telescope, and we shall scarcely ever approach close enough to an
alignment with the source and the deflector, he remarked that there is a 
little chance to observe
lensing phenomena caused by stellar-mass lenses. Though several
authors (Eddington 1920, Chwolson 1924)
 discussed that the deflection of light in a gravitational field may
give rise to GL, only  Zwicky (1937a,b) expressed his vision clearly
that the observation of the GL phenomena would be a certainty.
The studies of GL remained in almost a dormant stage until
Klimov, Liebes and Refsdal independently re-opened this subject
with their pioneering work in 1960's . Klimov (1963) investigated
the lensing of galaxies by galaxies, whereas Liebes (1964) studied
lensing of stars by  stars and also stars by  globular clusters in our
galaxy. Refsdal (1964a,b) was the first to argue that the geometrical
optics can be used  in studying the properties
of point-mass gravitational lenses and the time delay  resulting from
the lensing phenomena.

A quasar is an ideal source for GL because of its high
luminosity, point-like appearance, prominent spectral features
and the large distance from the earth (high probability of deflectors
intervening the source and the observer).  The great vision of
Zwicky became true when Walsh, Carswell and Weymann discovered
the first example of the GL - the QSO 0957+561 A,B (Walsh et al. 1979). The two
images of a single QSO were separated by six $ arcseconds$. Both
images had similar optical spectra, a galaxy was found between
the images, the flux ratio between images was found to be the
same in the optical as well as in the radio  wavebands, and VLBI
observations showed detailed correspondence between various
knots in the two radio images (cf. Narasimha et al. 1984, Narayan and
Bartelmann 1996). 
The VLBI features were modelled by Narasimha et al. (1984) who demonstrated
that a cluster in addition to the giant elliptical was needed for producing
the observed features.
 Following this
landmark discovery, more than a dozen confirmed examples of multiply-
imaged quasars are known and, in addition, six ring-shaped radio
images have been found (Refsdal and Surdej 1994, Keeton and Kochanek 1995). 
The Einstein rings
for extended sources were predicted by Saslaw et al. (1985) and were first
discovered by  Hewitt et al. (1987).
The detection of new lensing phenomena, namely giant luminous
arcs and arclets, and microlensing events have increased dramatically
in recent years. For a review on giant luminous arcs and arclets see
Fort and Miller(1994)  and on
microlensing see Narasimha (1995) and Roulet and Mollerch (1997).
 Cheng and Refsdal (1979, 1984)
as well as Subramanian et al. (1985) developed the theory
of microlensing to explain the flux variability in the images.
The traditional dynamical method to obtain masses of celestial
bodies in the universe requires that the system under investigation
is in dynamical equilibrium. Therefore, this method has limitation
in its use.  However, there are no such restrictions to the use
of GL and therefore, in recent years, it has become the most
important tool for probing the universe.
Indeed Narasimha and Chitre (1989) predicted that dark extended massive
objects could act as lenses for systems like 2345+007, MG 2016+112 etc.
which was confirmed with the discovery of a dark but X-ray bright cluster
at redshift of 1 in the system MG 2016+112 by Hattori et al. (1997).
Narasimha (1994) discussed the gravitational lens as a probe of dark
matter in the universe.
 The GL can give
important informations on masses of galaxies, the composition of
dark matter, cosmological parameters, existence of exotic massive
objects, the large scale structure of the universe and can be
also used to test the alternative theories of gravitation. A
gravitational lens provides  a highly magnified view of the distant
objects in the universe and therefore acts as a cosmic telescope.
Since the last decade the GL has become one of the most
important research fields in cosmology  (Schneider et al. 1992,
Refsdal and  Surdej 1994,  Narayan and Bartelmann 1996, Wu 1996).

The GL is very likely to serve an important tool to detect exotic objects in 
the universe, such as  cosmic strings and there have been attempts to
investigate if any of the confirmed gravitational lenses could be due to 
the action of  cosmic strings (see 
Hogan and Narayan 1984, Vilenkin 1984, 1986, Gott 1985, de Laix and Vachaspati
1996  and references therein). There is no compelling evidence
 that any of the observed gravitational lenses are due to a cosmic string.
However, it is essential to develop new gravitational lens models with objects
which are though not  yet observed,  are not forbidden on theoretical 
grounds.
It is well-known that the  scalar fields have been conjectured (since before 
the outset of the general theory of relativity) to give rise to the long-range
gravitational fields, and several theories involving scalar fields have been 
proposed (see 
Abraham 1914, Bergmann 1956, Brans and Dicke 1961, Callan et al. 1970, 
Garfinkle et al. 1991, 1992, Yilmaz 1992, Horne and Horowitz 1992, 1993
and references therein).  The  scalar fields  
minimally as well as conformally coupled to gravitation have been a subject of
active theoretical  research. In recent years, there is a growing interest in 
the studies of scalar (dilaton) fields, because of  their importance in string 
theories. 
There have been a number of studies of light propagation in the unconventional
scalar-tensor theories of gravity (Sanders 1989, Bekenstein and Sanders 1994
and references therein). However, the problem of formation of critical curves
and consequent appearance of multiple images  has not received adequate 
attention. 
In the present paper we study the effects of the massless scalar field on the
Schwarzschild lensing. More precisely, we build a new static and circularly
symmetric singular lens model characterized by two parameters: the Schwarzschild
mass and the ``scalar charge''. For this purpose we take the most general 
static spherically symmetric asymptotically flat exact solution to the 
Einstein-Massless Scalar (EMS) equations, given by Janis, Newman and Winicour 
(JNW) (Janis et al. 1968). Wyman (1981) obtained a static spherically symmetric
exact solution  to the EMS equations and later
Roberts (1993) showed that the most general static spherically symmetric
solution to the EMS equations (with zero cosmological constant)
is asymptotically flat and this is the Wyman solution. The Wyman solution is
well considered in the literature.
Recently one of us (Virbhadra 1997) showed that the Wyman solution is the same as 
the JNW solution, which was obtained about thirteen years ago, we 
therefore call it the JNW solution.
Switching off the ``scalar charge'' in this solution one 
recovers  the well-known Schwarzschild solution. Hereafter, we will refer to the 
new lens model as the Schwarzschild-Massless Scalar (SMS) lens. For  small
 values of the ratio
of the ``scalar charge'' to the Schwarzschild mass, the SMS lensing does not 
have any new  qualitative (though differs quantitatively) features and
resembles  the  Schwarzschild  lensing phenomena.
However, for  reasonably large value of this ratio, the
Einstein deflection angle starts with a negative value, becomes
positive, reaches a maximum and then decreases to zero as the
closest distance of approach increases from a small value to the
infinity.
This  behaviour of the SMS lens gives rise to few interesting
features, for instance, it gives radial critical curves
(radial caustics in the source plane) and concentric Einstein
double rings (tangential caustics in the source plane).  These
effects are not found in any of the known singular lens models. Moreover,
the number of images also differs from the case of the
Schwarzschild lensing.
 It is worth
mentioning that the JNW solution has a strong curvature globally
naked singularity (Virbhadra et al. 1997).  It is not known  how
one would be able to distinguish observationally  black holes from
 naked singularities (if these exist). Our present investigations 
reveal the qualitative different features and consequently the results
we present here could be useful in differentiating between the two kinds
of objects.
The rest of the paper is organized as follows: In 
 Sect. 2, we derive the Einstein deflection angle for a general
static and spherically symmetric metric, which generalizes the result
obtained by Weinberg (1972).   Sect. 3  discusses the JNW solution
for  obtaining  the deflection angle for this case.
In Sect. 4 we obtain positions of images, magnification, and
tangential and radial critical curves and in 
Sect. 5 we discuss possible observational tests. In the last Section
we summarize the qualitative features of the lens.
We  follow the  convention that Latin   indices take values $0\ldots3$ and  
use  geometrized units, for instance $ M \equiv G M/c^2$. 


%

\section{Einstein deflection angle for a general static
spherically symmetric metric }

We consider a general static and  spherically  symmetric
spacetime given by the  line element
\be
ds^2 = B\lt(r\rt) dt^2 - A\lt(r\rt) dr^2 -D\lt(r\rt) r^2 \lt(d\vartheta^2+\sin^2\vartheta\ d\phi^2\rt) .
\label{eq1}
\ee
The null geodesics equations are
\be
\frac{dv^i}{dk} + \Gamma^i_{jk} v^j v^k = 0,
\label{eq2}
\ee
where
\be
g_{ij} v^i v^j = 0.
\label{eq3}
\ee
$v^i \equiv \frac{dx^i}{dk}$ is the tangent vector to the null geodesics. $k$
is the affine parameter.

Eqs. $(2)$ with Eq.$(1)$ give
\be
B \frac{dt}{dk} = K ,
\label{eq4}
\ee
\be
D r^2 \sin^2\vartheta \frac{d\phi}{dk} = J ,
\label{eq5}
\ee
\be
\frac{d^2\vartheta}{dk^2}+\lt(\frac{2}{r}+\frac{D'}{D}\rt) \frac{dr}{dk}
\frac{d\vartheta}{dk} -\sin\vartheta
\cos\vartheta {\lt(  \frac{d\phi}{dk}   \rt)}^2 = 0 ,
\label{eq6}
\ee
and
\be
\frac{d^2r}{dk^2}+\frac{A'}{2A} \lt(\frac{dr}{dk}\rt)^2-\frac{D'r^2+2Dr}{2A} 
\lt[\lt( \frac{d\vartheta}{dk} \rt)^2+\sin^2\vartheta \lt(\frac{d\phi}{dk} \rt)^2\rt] 
+\frac{B'}{2A} \lt(\frac{dt}{dk}\rt)^2  = 0 .     
\label{eq7}
\ee
$K$ and $J$ are constants of integrations. The prime denotes the
derivative with respect to the  coordinate $r$. 
Without loss of generality we take
\be
 \frac{dt}{dk} = \frac{1}{B} .
\label{eq8}
\ee 
Appealing  to the spherically symmetric nature of the metric under consideration
we  consider the geodesics, without loss of generality, on
the equatorial plane ($\vartheta=\pi/2$). 
Following Weinberg (1972, Chapters 8.4 and 8.5), we get the equation
for the photon trajectories as 
\be
\phi(r)-\phi_{\infty} = {\int_r}^{\infty}
\lt(\frac{A(r)}{D(r)}\rt)^{1/2}
\lt[
\lt(\frac{r}{r_0}\rt)^2
\frac{D(r)}{D(r_0)} \frac{B(r_0)}{B(r)} -1
\rt]^{-1/2} \ \frac{dr}{r} ,
\label{eq9}
\ee
where $r_0$ is the closest distance of approach.
The integration constant $J$ is
\be
J = r_o  \sqrt{\frac{D(r_o)}{B(r_o)}} .
\label{eq10}
\ee 
The Einstein deflection angle is given by
\be
\hat{\alpha}\lt(r_0\rt)
 = 2 | \phi\lt(r_0\rt) - \phi_{\infty} | - \pi ,
\label{eq11}
\ee
where $\phi\lt(r_0\rt) - \phi_{\infty}$ is read through the Eq. $(9)$.
The above expression reduces to the result obtained by Weinberg (1972)
if one takes $D=1$.

\section{Janis-Newman-Winicour  solution and the deflection angle}

There have been  some studies of bending of light in scalar-tensor
theories (Bekenstein and Sanders 1994).
We wish to consider a gravitational lens endowed with the conventional
mass and a ``scalar charge'', described by the JNW solution.
We now  discuss the JNW solution in order to derive the Einstein deflection 
angle for the JNW metric. The Einstein-Massless Scalar equations (EMS) are
\be
R_{ij}\ -\ \frac{1}{2}\ R \ g_{ij}\ =\ 8 \pi \ S_{ij}\ ,
\ee
where $S_{ij}$, the energy-momentum tensor of the massless scalar field, is
given by
\be
S_{ij}\ =\ \Phi_{,i}\ \Phi_{,j}\ -\ \frac{1}{2}\ g_{ij}\ g^{ab}\ \Phi_{,a}\
          \Phi_{,b}\ ,
\ee
and 
\be
\Phi_{,i}^{\ ;i}\ =\ 0 .
\ee
 $\Phi$ stands for the massless scalar field. The comma and semicolon before 
indices denote  the partial and covariant derivatives, respectively.
$R_{ij}$ is the Ricci tensor and $R$ is the Ricci scalar.
Eq. $(12)$ with Eq. $(13)$ can be expressed as
\be
R_{ij}\ =\ 8 \pi \ \Phi_{,i}\ \Phi_{,j} .
\ee
JNW  obtained the most general   static  spherically symmetric 
asymptotically flat  exact solution to the EMS equations, which is given by
the line element (cf.  Virbhadra 1997)
\be
ds^2 = \lt(1-\frac{b}{r}\rt)^{\gamma} dt^2 
      - \lt(1-\frac{b}{r}\rt)^{-\gamma} dr^2
      - \lt(1-\frac{b}{r}\rt)^{1-\gamma} r^2
       \lt(d\vartheta^2\ +\ \sin^2\vartheta\ d\phi^2\rt)
\ee
and the scalar field
\be
\Phi = \frac{q}{b\sqrt{4\pi}} \ln\lt(1-\frac{b}{r}\rt),
\ee
where
\bea
\gamma &=& \frac{2M}{b}, \nn\\
b &=& 2 \sqrt{M^2+q^2}.
\eea
Note that $b < r< \infty$, and  $b$ is the curvature singularity; $M$ and $q$
are constant parameters which represent the total mass and the ``scalar charge''
respectively.
Clearly  $q=0$ recovers  the Schwarzschild solution.  The ``scalar charge''
 does not
contribute to the total mass of the system, but it does affect the
curvature of the spacetime (Virbhadra 1997). The JNW solution can be used in two
ways:  first, it describes the exterior gravitational field of 
an  object whose radius is greater than the parameter $b$ in
the solution. Second, it describes the field due a naked singularity (Virbhadra
et al. 1997).
Comparing Eq. $(1)$ with Eq. $(16)$, one has for the JNW solution
\bea
B &=& A^{-1} = \lt(1-\frac{b}{r}\rt)^{\gamma} , \nn\\
D &=& \lt(1-\frac{b}{r}\rt)^{1-\gamma} .
\eea
Substituting the above in Eq. $(11)$, one gets the Einstein deflection
angle
\be
\hat{\alpha}\lt(r_0\rt) = 2 \  {\int_{r_0}}^{\infty}
 \frac{dr}{r \sqrt{1-\frac{b}{r}} \  \sqrt{\lt(\frac{r}{r_0}\rt)^2 
\lt(1-\frac{b}{r}\rt)^{1-2\gamma} \lt(1-\frac{b}{r_0}\rt)^{2\gamma-1}
-1}} - \pi .
\ee
Introducing  dimensionless parameters 
\bea
x = \frac{r}{b} , \nn\\
x_0 = \frac{r_0}{b} , 
\eea
one can re-write the expression for the deflection angle
\be
\hat{\alpha}\lt(x_0\rt) = 2 \  {\int_{x_0}}^{\infty}
 \frac{dx}{x \sqrt{1-\frac{1}{x}} \  \sqrt{\lt(\frac{x}{x_0}\rt)^2 
\lt(1-\frac{1}{x}\rt)^{1-2\gamma} \lt(1-\frac{1}{x_0}\rt)^{2\gamma-1}
-1}} - \pi
\ee
The  integral in the above  is defined for $x_0 > (2 \gamma +1)/2$.
For the Schwarzschild metric ($\gamma = 1$)  this gives $r_0 > 3 M$.
The first derivative of the deflection angle is given by
\be
\hat{\alpha}'\lt(x_0\rt) = \frac{2\gamma+1-2x_0}{{x_0}^2\lt(1-\frac{1}{x_0}\rt)}
{\int_{x_0}}^{\infty}
 \frac{\lt(4 \gamma x - 2 \gamma - 1\rt) dx}
{\lt(2\gamma +1 - 2 x\rt)^2 \  x \ \sqrt{1-\frac{1}{x}} \  \sqrt{\lt(\frac{x}{x_0}\rt)^2 
\lt(1-\frac{1}{x}\rt)^{1-2\gamma} \lt(1-\frac{1}{x_0}\rt)^{2\gamma-1}
-1}} .
\ee
After lengthy, but straightforward calculations Eq. ($20$) gives the Einstein
deflection angle (up to the second order) as follows.
\be
\hat{\alpha}\lt(r_0\rt) =
 \frac{4M}{r_0} + \frac{4M^2}{{r_0}^2}\lt(\frac{15\pi}{16}-2\rt)
+\frac{2}{{r_0}^2}\lt[2M \sqrt{M^2+q^2} - \frac{q^2 \pi}{8}\rt]
+ . . .  . . \ \ \ .
\ee
We trivially recover the deflection angle for the Schwarzschild lens
when $q=0$ in the above equation. The second order term for the
Schwarzschild case is obviously positive.
In the following section we use equations $(22)$ and $(23)$ to perform
computations.
\section{ Image positions, magnification and critical curves }

We have given the  lens diagrams in Fig. 1. The first one is 
applicable to the case when the bending angle is positive  and the other one
is for the negative deflection.
We take the reference axis to be the line from  the observer O to the lens L
and keep the distance from the deflector to the
observer, $D_{ol}$, fixed. An image position is
specified by the angle  $\theta$ between OL and the tangent to the 
null geodesic at the observer. The observer is  assumed to be located in
an asymptotically  flat spacetime.
A null geodesic through the
observer is uniquely specified by the angle $\theta$.  The impact parameter
is $D_{ol}  \sin\theta$ and 
$D_{ls}$ stands for the distance from the source S to the point C in the lens 
plane. $D_{os}$ stands for the distance from  the observer to the source.
$\beta$ denotes  the true angular position
of the source, whereas $\theta$ stands for the image positions.
With these definitions, the lens equation may be expressed as
\be
\sin\lt(\theta-\beta\rt) = \frac{D_{ls}}{D_{os}} \sin\hat{\alpha} .
\ee
   \begin{figure*}
   \epsfbox{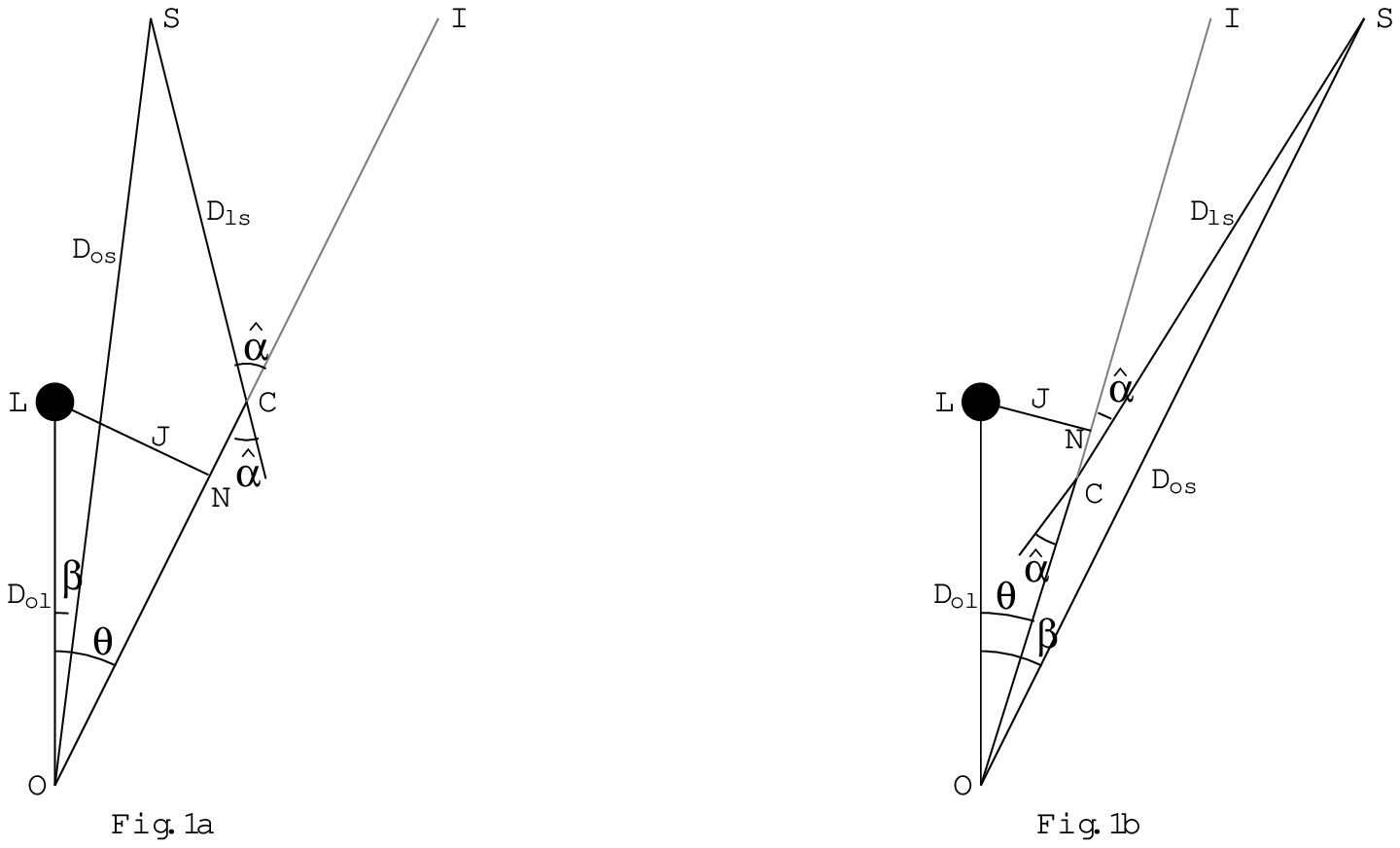}
 \caption[ ]{
The geometry of a gravitational lens: $S$ is the source location, $O$ is the
observer and $L$ is the lens. The angular separations of the source and the
image positions from the lens are denoted by $\beta$ and $\theta$, 
respectively. $\hat{\alpha}$ is the Einstein bending angle. The bending
angle is taken to be positive in case ${\bf a}$ where the contribution
due to the mass is dominant while it is negative in case ${\bf b}$ where
the ``scalar charge'' contributes more. $D_{os}$ represents the 
observer-source distance, $D_{ol}$ the observer-lens distance and $D_{ls}$
the lens-source distance. The lines $SC$ and $OI$ are tangents on the null 
geodesics at $S$ and $O$, respectively.
}
\label{lensdiag.eps}
\end{figure*}

We define
\be
\alpha = \sin^{-1}\lt(\frac{D_{ls}}{D_{os}} \sin\hat{\alpha}\rt) .
\ee

From the lens diagram one has
\be
\sin\theta = \frac{J}{D_{ol}} .
\ee
Using Eqs. $(10), (19), (21),$  and $(27)$  we get

\be
\sin\theta = \frac{b}{D_{ol}} \ 
          x_o \ \lt(1-\frac{1}{x_o}\rt)^{\frac{1-2\gamma}{2}} ,
\ee
where $b=2q$ for $\gamma=0$ and it is $2M/\gamma$ for $\gamma
\neq 0$.

The magnification of images is given by
\be
\mu = \lt( \frac{\sin{\beta}}{\sin{\theta}} \ \frac{d\beta}{d\theta} \rt)^{-1}.
\ee
The tangential and  radial critical curves  follow from the singularities
in
\be
\mu_t \equiv \lt(\frac{\sin{\beta}}{\sin{\theta}}\rt)^{-1} ,
\ee
and
\be
\mu_r \equiv \lt(\frac{d\beta}{d\theta}\rt)^{-1},
\ee
respectively.
To obtain magnification the first derivative of the deflection angle with 
respect to $\theta$ is required, which is given by
\be
\frac{d\hat{\alpha}}{d\theta} =  \hat{\alpha}'\lt(x_o\rt)
                               \frac{dx_o}{d\theta} ,
\ee
where
\be
 \frac{dx_o}{d\theta} = 
\frac{
   2 x_o \lt(1-\frac{1}{x_o}\rt)^{\frac{2\gamma+1}{2}}
   \sqrt{1-\lt(\frac{b}{D_{ol}}\rt)^2 {x_o}^2 \lt(1-\frac{1}{x_o}\rt)^{1-2\gamma}}
    }
{\frac{b}{D_{ol}} \lt(2x_o-2\gamma-1\rt)}.
\ee
\begin{figure*}
\epsfbox{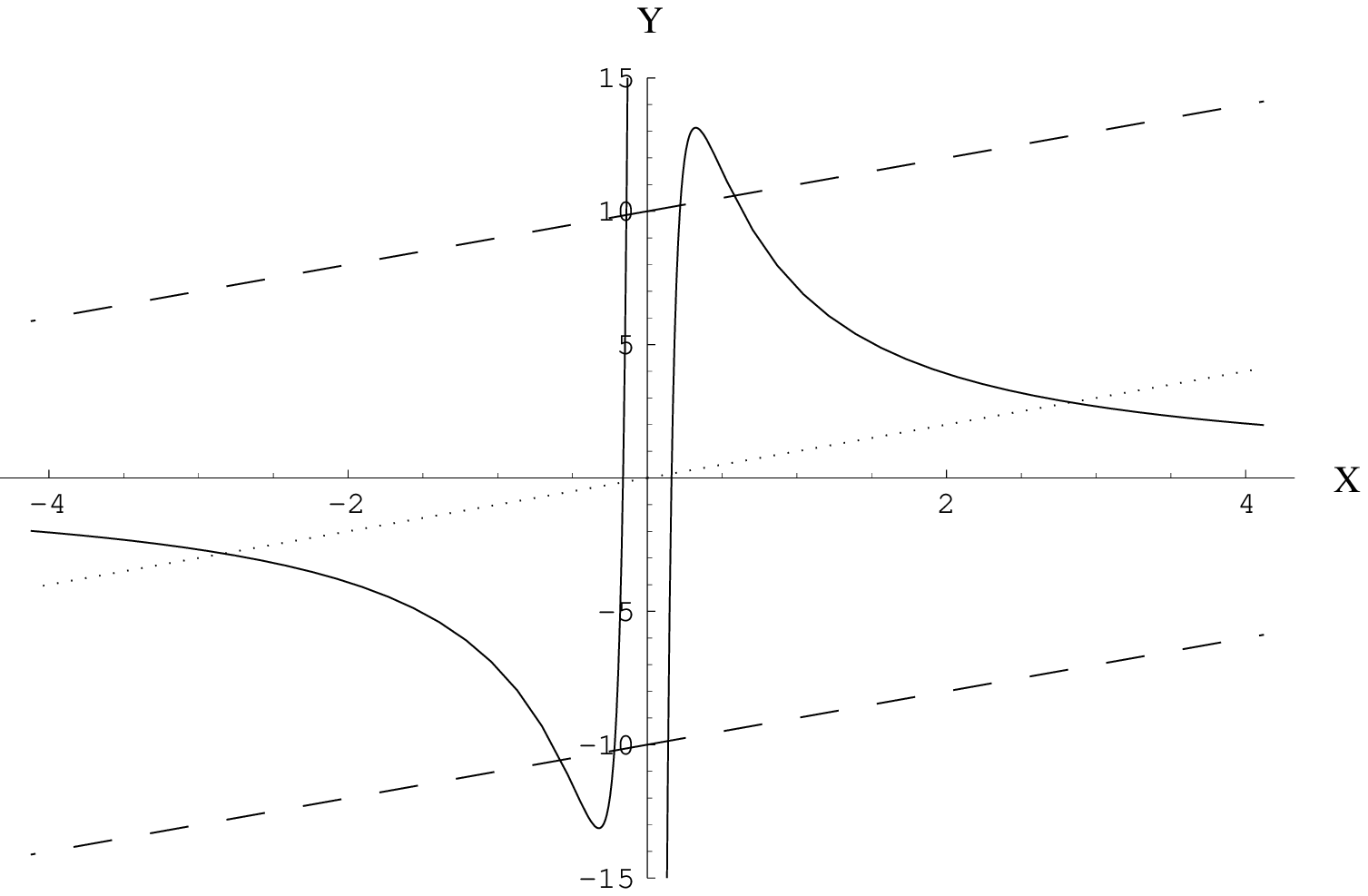}
\caption[ ]{
The deflection angle $\alpha$ and the angle $\theta-\beta$ are plotted as
a function of the angular separation $\theta$ of the image from the lens,
indicated by the continuous and dashed lines, respectively. The dotted 
line passing through the origin represents the source position precisely
aligned with the lens ($\beta =0$) and its intersections with the continuous
curves give the two Einstein radii.  For a specified position of the 
source the points of intersection of the continuous curves with  the
dashed line give the image positions. Here $\gamma = 0.005$, 
$D_{ol} = 10^{10} M$ and $\beta = \pm 10$. Angles are in arcseconds.
}
\label{images.eps}
\end{figure*}
\begin{table*}
 \caption[ ]{Image positions for $\gamma = 0.005$ }
 \begin{flushleft}
 \begin{tabular}{ccccc}
 \hline\noalign{\smallskip}
 \multicolumn{2}{c}{Images on the opposite side of the source}&
             Source position     &
 \multicolumn{2}{c}{Images on the same side of the source} \\
\qquad$\theta_{outer}$&$\theta_{inner}$&
$\beta$ &\qquad$\theta_{inner}$&$\theta_{outer}$ \\
 \hline\noalign{\smallskip}
\qquad\phantom{11}$2.5852$ &\phantom{11} $0.1641$ 
                &\qquad $0.5$ 
                &\qquad $0.1610$ &\phantom{11} $3.1008$ \\
\qquad\phantom{11}$2.3592$ &\phantom{11} $0.1658$ 
                &\qquad $1.0$ 
                &\qquad $0.1595$ &\phantom{11} $3.3901$ \\
\qquad\phantom{11}$1.8006$ &\phantom{11} $0.1712$ 
                &\qquad $2.5$ 
                &\qquad $0.1554$ &\phantom{11} $4.3735$ \\
\qquad\phantom{11}$1.1876$ &\phantom{11} $0.1822$ 
                &\qquad $5.0$ 
                &\qquad $0.1493$ &\phantom{11} $6.3132$ \\
\qquad\phantom{11}$0.8208$ &\phantom{11} $0.1972$ 
                &\qquad $7.5$ 
                &\qquad $0.1441$ &\phantom{11} $8.4838$ \\
\qquad\phantom{11}$0.5795$ &\phantom{11} $0.2203$ 
                &\qquad $10.0$ 
                &\qquad $0.1395$ &\phantom{11} $10.7776$ \\
\qquad\phantom{11}$0.3765$ &\phantom{11} $0.2783$ 
                &\qquad $12.5$ 
                &\qquad $0.1354$ &\phantom{11} $13.1396$ \\
\qquad\phantom{11}No image &\phantom{11} No image
                &\qquad $15.0$ 
                &\qquad $0.1317$ &\phantom{11} $15.5418$ \\
\qquad\phantom{11}No image &\phantom{11} No image 
                &\qquad $17.5$ 
                &\qquad $0.1284$ &\phantom{11} $17.9693$ \\
\qquad\phantom{11}No image &\phantom{11} No image
                &\qquad $20.0$ 
                &\qquad $0.1253$ &\phantom{11} $20.4135$ \\
 \noalign{\smallskip}
 \hline
 \noalign{\smallskip}
 \noalign{\smallskip}
 \end{tabular}
 \end{flushleft}
\begin{list}{}{}
\item[$^{\rm a}$] We have taken $D_{ol} = 10^{10} M$. Angles are in arcseconds.
\end{list}
 \end{table*}
\begin{figure*}
\epsfbox{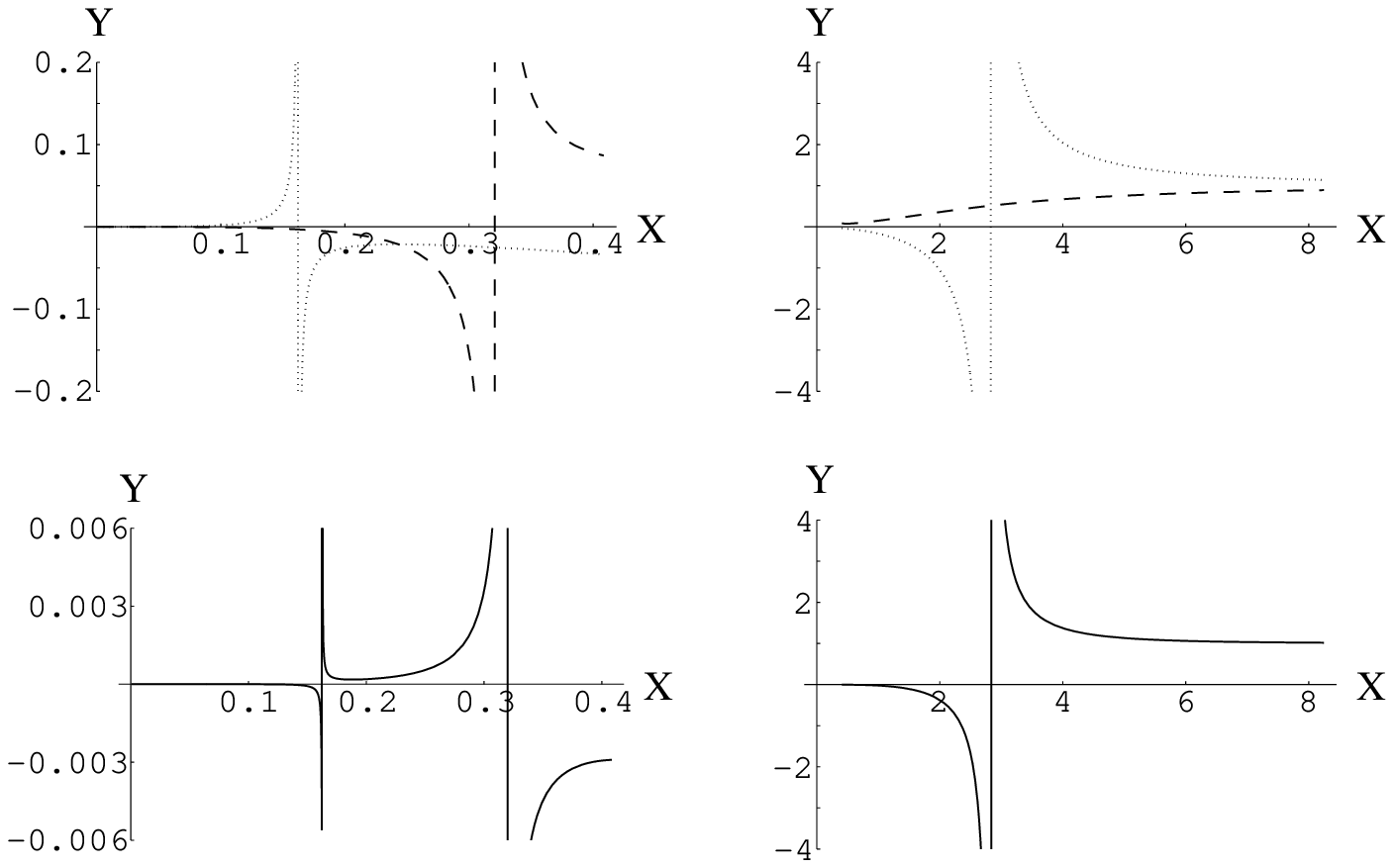}
\caption[ ]{
The {\em magnifications: tangential} $\mu_t$ denoted by dotted curves,
{\em radial}  $\mu_r$ denoted by dashed curves, and {\em total} $\mu$
shown as continuous curves are plotted as a function of the image position
$\theta$ for $\gamma=0.005$.  The singularities of $\mu_t$ and $\mu_r$
give positions of the tangential and radial critical curves, respectively.
Here $D_{ol} =  10^{10} M$ and angles are in arcseconds.
}
\end{figure*}
 
To obtain image positions on both sides of the optic axis one plots 
$\alpha\lt(\theta\rt), \theta-\beta$ against $\theta$ and 
$-\alpha\lt(\theta\rt), -\theta-\beta$ against $-\theta$. The points
of intersection give the image positions.  The continuous line on the
right (left) hand side of the vertical axis is $\alpha\lt(\theta\rt)$
against $\theta$ (-$\alpha\lt(\theta\rt)$ against $-\theta$). The 
dashed lines on the right (left)  hand side of the vertical
axis are $\theta-\beta$ against $\theta$ ($-\theta-\beta$ against 
$-\theta$). The dotted lines are, similarly, $\theta$ vs. $\theta$ and
$-\theta$ vs. $-\theta$. The intersections of $\alpha\lt(\theta\rt)$ with
$\theta$ give the angular positions of tangential critical curves (Einstein rings). For these
computations we have considered $D_{ls}/D_{os} = 1/2$ and $ D_{ol} = 10^{10} M $  
(for $\gamma \neq 0$). 
We have plotted for $\gamma = 0.005$  (see  Fig. 2). In Table 1, we have given
image positions for few
values of the source positions for $\gamma = 0.005$. In Fig. 3, we have
plotted tangential magnification $\mu_t$ and radial magnification $\mu_r$
against $\theta$ for $\gamma=0.005$.
Singularities in these give the positions of tangential critical curves (TCC)
and radial critical curves (RCC), respectively. In the same 
figure we have also plotted total magnification  $\mu$ vs. $\theta$. 
In Table 2, we give the positions of TCC and RCC for few values of $\gamma$.

For $\gamma = 0.005$, as shown in Fig. 2, $\alpha$ has large negative 
value for small $\theta$, becomes positive and then goes to zero (asymptotically) 
as $\theta$ increases.  The
dashed lines are for $\beta = \mp 10$ arcseconds. The dotted line ($\beta =0$)
cuts the continuous lines at two points on each side of the vertical axis
and therefore one gets two concentric Einstein rings. As $\beta$ increases
from zero (see the lower part of the figure) the dashed line cuts the
continuous curves at four points (two on each side of the vertical axis) giving
rise to four images (two images on each side of the optic axis). As $\beta$
increases the two images on the opposite side of the source come closer and
eventually meet at the RCC whereas those on the same side of the source get separated.
For any further increase in $\beta$ there is no image on the opposite side 
of the source. In Table 1, we have  given image positions for few
values of the source positions for $\gamma = 0.005$.
Out of the two images on the same side of the optic axis, the one which is
nearer to the optic axis we call the {\em inner image} and the other which
is farther from the optic axis we call the {\em outer image}.
 For this particular case we  have plotted $\mu_t$ and $\mu_r$ to
identify the TCC and RCC (see Fig. 3). Magnification $\mu$ is
also plotted.  We have two TCC and one RCC (the RCC lies in
between the two TCC). The magnification falls off sharply near the inner
TCC as compared to the RCC as well as the outer TCC.
For $\gamma = 0.001$, $\alpha$ has large negative value for small $\theta$,
becomes positive and then goes to zero (asymptotically) as $\theta$ increases.
There are two images of opposite parities
on  the same side of the source. As $|\beta|$ decreases the two images
meet at the RCC and for any further decrease in $|\beta|$ there is no
image. There is no  Einstein ring for this case.
For $\gamma \geq 0.5$ the GL is qualitatively similar to the Schwarzschild lensing
(one has only one TCC and has two images, one on each side of the optic axis).

We have given  the angular positions of critical curves for
few values of $\gamma$. $\theta_t^{inner}, \theta_r$ and
$\theta_t^{outer}$ stand for the angular positions of the inner
TCC, RCC, and outer TCC, respectively.
\begin{table*}
\caption[ ]{Positions of  critical curves}
\begin{flushleft}
\begin{tabular}{lllll}
\hline
$\gamma$ &   $\Theta_t^{(inner)}$ &  $\Theta_r$ &  $\Theta_t^{(outer)}$ \\
\noalign{\smallskip}
\noalign{\smallskip}
\hline\noalign{\smallskip}
$0.001$ &  No      & $3.4159$  &  No     \\
$0.002$ & $1.2326$ & $1.5701$  & $2.0984$\\
$0.003$ & $0.4616$ & $0.8323$  & $2.6588$\\
$0.004$ & $0.2551$ & $0.4923$  & $2.7812$\\
$0.005$ & $0.1625$ & $0.3202$  & $2.8324$\\
$0.01 $ & $0.0405$ & $0.0810$  & $2.8966$\\
$0.5  $ &  No      &    No     & $2.9170$\\
$1.0  $ &  No      &    No     & $2.9171$\\
\noalign{\smallskip}
\hline
\end{tabular}
\end{flushleft}
\begin{list}{}{}
\item[{}]
$\theta_t^{inner}, \theta_r$ and
$\theta_t^{outer}$ stand for the angular positions of the inner
tangential curves, radial critical curves, and  outer tangential curves
, respectively.  We have taken $D_{ol} = 10^{10} M$ and angles are 
expressed in arcseconds.
\end{list}
\end{table*}
An increase in the ``scalar charge'' to the mass ratio (decrease in $\gamma$)
decreases the value of angular position of the outer TCC and increases the
value of the inner TCC and the RCC. The angular separation between the 
two Einstein rings decreases with the increase in this ratio. For ``very
large''
value of this ratio (for example, for $\gamma \leq 0.001$) there is only RCC and
there is no TCC and for the small value of this ratio (for instance, for
$\gamma \geq 0.5$)
there is only one TCC  and no RCC. For ``large'' value of this
ratio (for instance, for $\gamma = 0.005, 0.01$)
there is one RCC, which lies in between two TCC's. 
In the entire investigations we have not considered points close to the
integral singularities.
\section{Gravitational lens as a diagnostic of scalar charge}

We have given, in Table 1, typical solutions for the positions of images 
due to lensing by ``scalar charge'' and normal mass and have shown the same 
in Fig. 2. The corresponding critical curves are given in Table 2 and are
displayed in Fig. 3.  
The observations of double Einstein rings near the
centres of relativistic star clusters in galactic nuclei 
could be a powerful probe of ``scalar charge''.
The most important property of such a configuration useful for the
diagnostic of the lens is the formation of four images of a background
radio source located at about an arcsecond from the line of sight to the
centre of the lens. Two of the images will be  at opposite sides of the lens
and will be bright, but the other two images (one on each side of the lens)
will be generally faint.
 For a source sufficiently away
 from the optic axis, two images will be formed, {\it both
in the same side of the lens} unlike the normal mass distribution acting
as lens. The present VLBA and VLBI are
capable of detecting the radio images at milliarcseconds and inferring
from the spectral characteristics that they could indeed be images of the
same background object.  The ellipticity of the
mass distribution in the lens could alter the image configuration. Still,
if we do indeed detect four radio images nearly collinear, it will lend 
support to possible existence of ``scalar charge''.
 More probable
configuration will, of course, be formation of two images on the same side
of the lens. Though difficult to detect, if these features
should turn out to be present, it could provide a means to determine the
relative gravitational influence of the ``scalar charge'' q and  mass M in
the lens.

Our results become important if the following  observations yield
positive results:
Observations  of relativistic star clusters 
in galactic nuclei  should show evidence for  double ring structure or four
images of a background source along a line or two images in the same side
of the lens with no evidence for a third image in the other side.
These  results, if found to be true, will strongly argue in favour of
existence of ``scalar charges'' in the universe.

\section{Discussion and Summary}

The gravitational lens  can serve as a very useful tool to discover 
exotic objects in the universe
(described by matter fields which are not detected, as yet) as well as
to test alternative theories of gravity. 
Even  before the advent of the general theory of relativity, scalar fields
have been proposed to simulate the long-range gravitational fields and
it still continues (see in Horne and Horowitz 1992, Yilmaz 1992 and references
therein). Bekenstein and Sanders (1994) discussed that gravitational lenses
in scalar-tensor theories could be divergent under some circumstances.
We studied the null geodesics
in a general static spherically symmetric spacetime and calculated
the Einstein deflection angle.
The deflection angle for a general static spherically symmetric metric
obtained in this work can be used to construct  circular gravitational
lens models (with different matter fields) in the Einstein theory as well as in
alternative theories of gravity.   We have used this result to study
a circularly
symmetric gravitational lens in the Einstein-Massless Scalar theory,
by  considering  the most
general static and spherically symmetric solution, 
given by Janis et al. This describes the exterior field due a spherically
symmetric massive object endowed with ``scalar charge''. 
The same field also describes the spacetime due to a naked 
singularity. Though the existence of a naked singularity is debatable,
this subject has recently attracted the attention of  many researchers'
minds (see Virbhadra et al. 1997 and references therein). 
It is not altogether  clear how one would be
able to distinguish observationally a black hole from a naked singularity
(if indeed these exist).
We propose that the present  lens model could be helpful for this purpose.

The new Schwarzschild-Massless Scalar gravitational lens has  interesting 
features which are not
shared by other known circularly symmetric gravitational lenses. For ``small''
values of the ``scalar charge'' to the mass ratio  the
lensing is qualitatively similar (but  quantitatively different) to the 
Schwarzschild lens. For these cases, like the Schwarzschild lens, there
are two images of opposite parities (one on each side of the optic axis),
there exits no radial critical curve, and there is only the Einstein ring.
 The ``scalar charge'' contributes to the
decrease in the radius of the Einstein ring. However, when the ``scalar charge''
to the mass ratio is not small, the gravitational lens action  is quite 
different from the Schwarzschild lensing. When this ratio is ``very large''
(for example, $\gamma \leq 0.001$) one gets two 
images of opposite parities   on {\em the same side of the source}
or nil depending upon the source position.
 There exists one radial critical curve and no tangential
critical curve (no Einstein ring) in  this case.
However, for ``large'' ratio cases there are four images (two 
 on each side of the optic
axis) or two   on the same side of the source, depending upon the location
of  source. There exist two tangential critical curves ({\em double Einstein
 rings}) and
one radial critical curve which lies in  between them. One interesting feature 
of these
cases (``very large'' as well as ``large'' ratio) is that one can get 
images  of positive parity and sometimes large magnification at smaller angles
compared to the source position, cf. Table 1.
 An image which forms at smaller angular position compared
to its source position are usually demagnified, unless it is close to a
critical curve. Thus, under some situations one can get magnified image at 
smaller angular position compared to its source position.


\begin{acknowledgements}
Thanks are due to H. M. Antia and Kerri (R. P. A. C.) Newman for valuable
discussions.  Thanks are also due to 
J. M. Aguirregabiria,
M. Dominik,
J. Kormendy,
J. Lehar,
S. Shapiro 
and
J. Surdej
for helpful correspondence.
This research was partially  supported by FRD, S. Africa. 
One of us (KSV) thanks K. MacKay   for his   kind hospitality.
\end{acknowledgements}


\begin{thebibliography}{}

\bibitem{}  Abraham M., 1914, Jahrb. Radioakt. Electronik  11, 470 

\bibitem{}  Bekenstein J. D., Sanders  R. H., 1994, ApJ 429, 480

\bibitem{}  Bergmann O, 1956, Am. J. Phys.  24, 38 

\bibitem{}  Brans C.  H.,  Dicke R. H., 1961, Phys. Rev. 124, 925 

\bibitem{}  Callan C. G, Jr.,  Coleman S.,   Jackiw R., 1970,
            Ann. Phys. (N.Y.)  59, 42

\bibitem{}  Cheng K.,  Refsdal S., 1979, Nat  282, 561

\bibitem{}  Cheng K.,  Refsdal S., 1984,  A \& A 132, 168 

\bibitem{}  de  Laix A. A.,  Vachaspati T., 1996, Gravitational lensing
               by cosmic strings loops, astro-ph/9605171

\bibitem{}  Eddington A. S., 1920,   Space, time and gravitation,
                  Cambridge University  Press, Cambridge

\bibitem{}  Einstein A., 1936,  Sci  84,  506 

\bibitem{}  Fort B.,  Miller Y., 1994,  A\&AR  5, 239 

\bibitem{}  Garfinkle D,  Horowitz G.T.,   Strominger A., 1991,
            Phys. Rev.  D43, 3140

\bibitem{}  Garfinkle D.,  Horowitz G.T.,   Strominger A., 1992, 
            Erratum :  Phys. Rev.  D45, 3888  

\bibitem{}  Gott III J., R., 1985,  ApJ 228, 422 

\bibitem{} Hattori M. et al., 1997, Nat. 388, 146

\bibitem{}  Hewitt J. N. et al., 1987,  ApJ  321, 706 

\bibitem{}  Hogan C. J.,  Narayan R., 1984, MNRAS  2111, 575

\bibitem{}  Horne J. H.,  Horowitz G.T., 1992, Phys. Rev.  D46, 1340 

\bibitem{}  Horne  J. H.,  Horowitz G. T., 1993, Nucl. Phys.  B399, 169 

\bibitem{}  Janis A. I.,  Newman E. T.,   Winicour J., 1968,
            Phys. Rev. Lett.   20,  878 

\bibitem{}  Klimov Yu. G., 1963, Sov. Phys. Doklady  8, 119 

\bibitem{}  Keeton II C. R.,  Kochanek C. S., in: Astrophysical 
            Applications of Gravitational Lensing, IAU Symp., 173, 
           eds. C. S. Kochanek and  J. N. Hewitt, Kluver, Boston, p419

\bibitem{}  Liebes Jr. S., 1964,  Phys. Rev.  133, B835 

\bibitem{} Narasimha D., Subramanian K., Chitre S. M., 1984, MNRAS 210, 79 

\bibitem{} Narasimha D., Chitre S. M., 1989, Astron. J. 97, 327

\bibitem{} Narasimha D., 1994, in: ICNAPP, ed. R. Cowsik, 251

\bibitem{} Narasimha D., 1995, Bull. Astr. Soc. India  23, 489 

\bibitem{}  Narayan R., Bartelmann M., 1996,  Lectures on gravitational 
            lensing, astro-ph/9606001

\bibitem{}  Refsdal S., 1964a,  MNRAS 128, 295

\bibitem{}  Refsdal S., 1964b,  MNRAS 128, 307 

\bibitem{}  Refsdal S.,   Surdej J., 1994,  Rep. Prog. Theor. Phys. 57, 
            117 

\bibitem{}  Roberts M. D., 1993, Ap\&SS, 200, 331

\bibitem{}  Roulet E.,   Mollerch S., 1997,  Phys. Rep.  279, 67 

\bibitem{} Sanders, R. H., 1989, MNRAS 241, 135

\bibitem{}  Saslaw W C, Narasimha D., Chitre S. M., 1985, ApJ 292, 348

\bibitem{}  Schneider P.,  Ehlers J.,  Falco E. E., 1992, Gravitational 
            Lenses,   Springer Verlag, Berlin 

\bibitem{}  Subramaniam K., Chitre S. M.,   Narasimha D., 1985, ApJ 289, 37 
          
\bibitem{}  Vilenkin A., 1984,  ApJ  L51, 282 

\bibitem{}  Vilenkin A., 1986,  Nat.  322, 613 

\bibitem{}  Virbhadra K. S. ,  Jhingan S.,  Joshi P. S., 1997,
                       Int. J. Mod. Phys.  D6, 357

\bibitem{}  Virbhadra K. S., 1997,
                       Int. J. Mod. Phys.  A12, 4831

\bibitem{}  Walsh D., Carswell R. F.,   Weyman R.J., 1979  Nature 279, 
            381 

\bibitem{}  Weinberg S., 1972,  Gravitation and cosmology: principles  and
            applications  of the general  theory of relativity,
            John Wiely \& Sons, NY

\bibitem{}  Wyman M., 1981,  Phys. Rev.  D24, 839 

\bibitem{}  Wu X. P., 1996, Fund. Cosmic Phys.  17, 1  

\bibitem{}  Yilmaz H., 1992, Il Nuovo Cimento  107B, 941 

\bibitem{}  Zwicky F., 1937a,  Phys. Rev.  51, 290

\bibitem{}  Zwicky F., 1937b,  Phys. Rev.  51,  679


\end{thebibliography}
\end{document}